\documentclass[showpacs]{revtex4} 
\usepackage{epsfig,amsmath,amssymb,graphicx}
\usepackage{natbib}
\usepackage[english]{babel} 

\begin{document}

\vspace{0mm}
\title{ Waves in a Bose-Einstein condensate of atoms with a dipole moment} %
\author{Yu.M. Poluektov, A.A. Soroka}
\email{yuripoluektov@kipt.kharkov.ua} %
\affiliation{National Science Center ``Kharkov Institute of Physics
and Technology'', Akhiezer Institute for Theoretical Physics, 61108
Kharkov, Ukraine}

\begin{abstract}
Based on the modified Gross-Pitaevskii equation for atoms with
intrinsic dipole moments which accounts for the relaxation of a
condensate, the dipole-dipole interaction and the interaction of
atoms with the electromagnetic field, the propagation of the sound
and electromagnetic waves in a Bose-Einstein condensate is studied. %
Owing to hybridization of the electromagnetic and sound waves near
the resonance frequency of an atom, there arise the two branches of
excitations in which the electromagnetic oscillations transform into
the sound oscillations and vice versa. It is shown that under
hybridization the crucial role is played by the dipole-dipole
interaction, which leads to a substantial increase of the repulsion
between the branches of the spectrum. The influence of the
dissipative effects connected with the relaxation of the macroscopic
wave function of the condensate and the imaginary part of the
polarizability of an atom on the form of the dispersion curves is
investigated.
\newline%
\end{abstract}
\maketitle

\section{Introduction}\vspace{-0mm} 

Recently, considerable attention is paid to investigation of the
behavior of the superfluid systems of Bose particles in the electric
and magnetic fields. The interest in these problems is stimulated by
the experimental investigations of Bose-Einstein condensates (BEC)
in the traps created by both magnetic fields and the field of the
laser radiation [1,2]. A special character of the interaction at low
temperatures of the electromagnetic field with many-particle systems
of electrically neutral atoms obeying the Bose-Einstein statistics
is indicated by the investigations of the propagation of light in
atomic gases [3,4]. An unexpectedly high electrical activity of
superfluid helium, manifested in different conditions, was reported
in a series of experimental works [5-11]. Understanding of the
effects observed in such and similar experiments requires a detailed
investigation of the interaction of many-particle systems of Bose
atoms, being in the coherent state, with the electromagnetic field.

The interaction of the electromagnetic field with a system of
electric charges is realized through the multipole moments of a
system. If the system is electrically neutral, then the next in
importance characteristic that describes the interaction with the
electric field is its dipole moment. There are arguments in favor of
that a helium atom, which does not have an intrinsic  dipole moment
in the free state, can spontaneously acquire an intrinsic dipole
moment in the liquid helium due to the interaction with the
surrounding atoms [12,13]. In this respect, it appears important a
detailed theoretical study of the electrical properties of the
superfluid system of atoms possessing an intrinsic dipole moment.
This will allow to compare theoretical predictions with the
phenomena observed in experiments. The propagation of the
electromagnetic field in BEC with account of the inner structure of
atoms within the model of an ideal gas was studied in [14-16], and
the account for the structure of atoms within the modified
Gross-Pitaevskii approach (GP) was suggested in [12].

In this work, basing on the GP equation that accounts for the
relaxation in BEC, the interaction with the electromagnetic field
and the dipole-dipole interaction, the propagation of the
electromagnetic and sound waves in the anisotropic superfluid system
of atoms with intrinsic dipole moments is theoretically studied.
Both an intrinsic dipole moment and a dipole moment induced by the
electric field is taken into account. The permittivity of the
condensate of atoms is calculated. It is shown that a BEC of atoms
with dipole moments is a medium with the frequency and spatial
dispersion. Since particles possess intrinsic dipole moments, there
arises a relationship between the electromagnetic and sound waves in
such a medium, which is particularly enhanced at a frequency close
to the resonance frequency of an atom. The propagation of the sound
waves in the condensate is accompanied by the oscillations of the
intensity of the electric field, and while the electromagnetic wave
is propagating there arise the oscillations of the condensate. It
proves out that taking into account of the dipole-dipole interaction
is essential for the description of the hybridization effect. The
influence of the dissipative processes on the propagation of waves
in a BEC of particles with a dipole moment is considered.

\section{Gross-Pitaevskii equation for atoms with a dipole
moment in the electromagnetic field with account of the dissipation
and dipole-dipole interaction }\vspace{-0mm} %
The dynamical GP equation for the macroscopic wave function of the
condensate $\psi=\psi({\bf r},t)$, interacting with the alternating
external field $U({\bf r},t)$ [1,2], can be obtained by means of the
Lagrange formalism [17,18], if the Lagrange function is chosen in
the form
\begin{equation} \label{01}
\begin{array}{l}
\displaystyle{%
  L=\int\!d{\bf r}\left[i\frac{\hbar}{2}\left(\psi^*({\bf r})\dot{\psi}({\bf r})-\dot{\psi}^*({\bf r})\psi({\bf r}) \right)
  -\frac{\hbar^2}{2m}\big|\nabla\psi({\bf r})\big|^2-U({\bf r},t)\big|\psi({\bf r})\big|^2 \right]
  -\frac{1}{2}\int\!d{\bf r}d{\bf r}'U({\bf r},{\bf r}')\big|\psi({\bf r})\big|^2\big|\psi({\bf r}')\big|^2. %
}
\end{array}
\end{equation}
For brevity the designation of dependence of the wave function on
time is omitted. The potential of the interparticle interaction can
be presented in the form
\begin{equation} \label{02}
\begin{array}{l}
\displaystyle{%
  U({\bf r},{\bf r}')=g\delta({\bf r}-{\bf r}')+U_D({\bf r},{\bf r}'), %
}
\end{array}
\end{equation}
where the first term accounts for the short-range interaction, and
the second term
\begin{equation} \label{03}
\begin{array}{l}
\displaystyle{%
  U_D({\bf r},{\bf r}')=\frac{{\bf d}({\bf r})\cdot {\bf d}({\bf r}')-3\big[{\bf x}\cdot {\bf d}({\bf r})\big]\big[{\bf x}\cdot {\bf d}({\bf r}')\big]}{R^3} %
}
\end{array}
\end{equation}
-- the long-range dipole-dipole interaction, and in (\ref{03}) ${\bf R}={\bf r}'-{\bf r}$, ${\bf x}\equiv{\bf R}/R$. %
The Lagrange function (\ref{01}) leads to the equation
\begin{equation} \label{04}
\begin{array}{l}
\displaystyle{%
  i\hbar\dot{\psi}= -\frac{\hbar^2}{2m}\triangle\psi+U({\bf r},t)\,\psi+\psi\int\! U({\bf r},{\bf r}')\big|\psi({\bf r}',t)\big|^2d{\bf r}'. %
}
\end{array}
\end{equation}
The equation (\ref{04}) is reversible in time, that is invariant
with respect to the transformation $t\rightarrow -t, \psi\rightarrow
\psi^*$, and describes the dynamics of the condensate without taking
into account of possible dissipative processes. In the absence of
the alternating external field the stationary solution of this
equation has the form $\displaystyle{\psi({\bf
r},t)=\widetilde{\psi}_0({\bf r})\exp\!\!\Big[\!-i\frac{\mu
t}{\hbar}\Big]}$. In the nonstationary case, in the presence of the
alternating external field, the function $\widetilde{\psi}_0({\bf
r},t)$ depends on time and the equation (\ref{04}) acquires the form
\begin{equation} \label{05}
\begin{array}{l}
\displaystyle{%
  i\hbar\dot{\widetilde{\psi}}= -\frac{\hbar^2}{2m}\triangle\widetilde{\psi}+\big[U({\bf r},t)-\mu\big]\widetilde{\psi}+\widetilde{\psi}\int\! U({\bf r},{\bf r}')\big|\widetilde{\psi}({\bf r}',t)\big|^2d{\bf r}', %
}
\end{array}
\end{equation}
where now $\widetilde{\psi}\equiv \widetilde{\psi}_0({\bf r},t)$. In
the following the macroscopic wave function is treated as precisely
a function which does not depend on time in the stationary state,
and the sign of tilde is omitted. The parameter $\mu$ has the
meaning of the chemical potential and in the equilibrium state is
related to the total number of particles by the equation
\begin{equation} \label{06}
\begin{array}{l}
\displaystyle{%
  N=\int\!\big|\psi_0({\bf r})\big|^2d{\bf r}.
}
\end{array}
\end{equation}

The equation (5) entails the conservation law for the total number
of particles in the condensate. Meanwhile, it is evident that under
nonstationary processes the particles from the condensate can pass
to the overcondensate quasiparticle states, so that the number of
particles in the condensate will be no longer conserved. If at some
moment of time the system resides in a nonstationary state, in which
a part of the particles is in the one-particle condensate and the
rest part of the particles forms a gas of quasiparticles, then such
state will relax to the equilibrium state, and at zero temperature
with time all the particles will pass into the condensate. The
damping of oscillations of the atomic condensate in magnetic traps
was observed experimentally [19] and turned out to be small.
Dissipative processes in the one-particle condensate can be taken
into account phenomenologically within the Lagrange formalism by
introducing the dissipative function [17], which is usually chosen
to be quadratic in the velocity [20], so that
\begin{equation} \label{07}
\begin{array}{l}
\displaystyle{%
  D=\hbar\gamma\dot{\psi}^*\dot{\psi},
}
\end{array}
\end{equation}
and $\gamma$ is a real phenomenological dimensionless dissipative
coefficient. As a result, in view of (\ref{01}) and (\ref{07}), we
come to the equation which differs from the equation (\ref{05}) only
by the presence of the time derivative with a real coefficient
\begin{equation} \label{08}
\begin{array}{l}
\displaystyle{%
  i\hbar\dot\psi= -\frac{\hbar^2}{2m}\triangle\psi+\big[U({\bf r},t)-\mu\big]\psi+\psi\int\! U({\bf r},{\bf r}')\big|\psi({\bf r}',t)\big|^2d{\bf r}'+\hbar\gamma\dot{\psi}. %
}
\end{array}
\end{equation}
Formally (\ref{08}) is obtained from (\ref{05}) by the substitution
$i\rightarrow i-\gamma$.

Let us make a remark as to accounting for the dissipative phenomena
in this approach. The standard GP equation is applied for the
description of the condensate at zero temperature. Dissipative
processes are accompanied with the production of entropy and the
heat release and, consequently, with the increase in temperature. In
the approach under consideration these temperature effects are
disregarded and it is supposed that, as in the standard case, the
system is considered at zero temperature. Actually it means that the
released heat is being removed from the system so fast that its
temperature does not have time to change noticeably. Such
approximate account of the dissipation is entirely equivalent to the
account of friction in mechanics, where the transformation of the
mechanical energy into heat is described by means of the dissipative
function, but the notions of temperature and entropy are not used in
this case.

It can be shown [21] that the introduced dimensionless dissipative
coefficient $\gamma$ is connected to the third viscosity coefficient
$\zeta_3$ and the time of uniform relaxation of the particle number
density in the condensate $\tau_0$ by the relations
\begin{equation} \label{09}
\begin{array}{l}
\displaystyle{%
   \zeta_3=\frac{\hbar\gamma}{2m^2n_0}, \qquad \tau_0=\frac{\hbar}{2g\gamma
   n_0}.
}
\end{array}
\end{equation}
Let us estimate the value of the coefficient $\gamma$  under the
assumption that relations (\ref{09}) are valid as well for the
superfluid $^4$He. For the helium third viscosity coefficient there
is an estimation [22]
\begin{equation} \label{10}
\begin{array}{l}
\displaystyle{%
   2\cdot 10^{-3}\,\, cm^5/g\!\cdot\!s \, < \, \zeta_3 \,  < \,  1.6\cdot 10^{-2}\,\, cm^5/g\!\cdot\!s. %
}
\end{array}
\end{equation}
From here and (\ref{09}) it follows that
\begin{equation} \label{11}
\begin{array}{l}
\displaystyle{%
   4< \gamma < 30. %
}
\end{array}
\end{equation}
In calculations for atomic condensates one often uses a less value
$\gamma\approx 0.01$ [23].

The potential energy of interaction of the condensate atom with the
electric field  in the dipole approximation has the form
\begin{equation} \label{12}
\begin{array}{l}
\displaystyle{%
   U({\bf r},t)=-{\bf d}\cdot {\bf E}, %
}
\end{array}
\end{equation}
where ${\bf d}$ is the dipole moment of an atom, ${\bf E}({\bf
r},t)={\bf E}_0+\delta{\bf E}({\bf r},t)$ is the electric field
acting on an atom, which is a sum of the constant and alternating
fields. As is known [24], the local electric field acting on an atom
in the dielectric differs from the external field, but for the
diluted systems such as atomic condensates this difference is very
small, it is small as well for the superfluid helium. Therefore we
consider ${\bf E}_0=E_0{\bf e}$ as the constant external field
directed along the unit vector {\bf e}. The alternating part of the
field arises during the propagation of the electromagnetic wave.

The dipole moment of an atom is a sum of two terms:
\begin{equation} \label{13}
\begin{array}{l}
\displaystyle{%
   {\bf d}={\bf d}_0+{\bf d}_p.
}
\end{array}
\end{equation}
Here ${\bf d}_0$ is the intrinsic dipole moment of an atom which is
directed along the external field in the equilibrium state. Its
value $d_0$ is considered constant. The second term is the dipole
moment induced by the field
\begin{equation} \label{14}
\begin{array}{l}
\displaystyle{%
   {\bf d}_p\equiv  {\bf d}_p(t)=\int_{-\infty}^t \alpha(t-t'){\bf
   E}(t')dt'=\int_{0}^\infty\!\alpha(\tau){\bf E}(t-\tau)d\tau, %
}
\end{array}
\end{equation}
where $\alpha$ is the polarizability of a single atom. The intrinsic
dipole moment, generally speaking, can change its direction, so that
in general case ${\bf d}_0(t)=d_0{\bf e}+\delta {\bf d}_0(t)$. Here
the first term is the equilibrium dipole moment and the second term
is the part depending on time. Since the value of the intrinsic
dipole moment is constant, then at small deviations from equilibrium
$\delta {\bf d}_0(t)\,{\bf e}=0$. As can be seen below, by virtue of
the last condition the fluctuations of the intrinsic dipole moment
fall out of the equations for the macroscopic wave function of the
condensate. Hence in this approximation at small oscillations the
intrinsic dipole moment can be considered constant both in magnitude
and direction, so that in the following we assume ${\bf d}_0=d_0{\bf e}$. %

The polarization part of the dipole moment equals ${\bf
d}_p(t)=\alpha_0E_0{\bf e}+\delta {\bf d}_p(t)$, where
$\alpha_0\equiv\alpha(0)$ is the static polarizability of an atom.
Here the first term is the constant dipole moment induced by the
constant field and the second term is the dipole moment induced by
the alternating field
\begin{equation} \label{15}
\begin{array}{l}
\displaystyle{%
   \delta{\bf d}_p(t)=\int_{0}^\infty\!\alpha(\tau)\delta{\bf E}(t-\tau)d\tau. %
}
\end{array}
\end{equation}
Thus, the total dipole moment of an atom can be presented as a sum
of the equilibrium and alternating parts ${\bf d}(t)=d_s{\bf
e}+\delta {\bf d}(t)$, where $d_s=d_0+\alpha_0 E_0$. The index of
the nonstationary dipole moment in (\ref{15}) will be omitted in the
following, so that $\delta {\bf d}(t)\equiv \delta {\bf d}_p(t)$.

In the stationary equilibrium state the phase of the order parameter
$\psi_0$ can be chosen so that it would be real, and from (\ref{08})
it follows
\begin{equation} \label{16}
\begin{array}{l}
\displaystyle{%
   \psi_0^2\equiv n_0=\frac{\mu+d_sE_0}{g^*},     %
}
\end{array}
\end{equation}
where $g^*\equiv\int U({\bf r},{\bf r}')d{\bf r}'$. In the spatially
uniform equilibrium state the dipole-dipole interaction does not
influence on the value of the equilibrium density, because its
contribution becomes zero under integration over angles, so that
$g^*=g$. In view of (\ref{16}), GP equation which accounts for the
relaxation, the interaction with the electric field in the dipole
approximation and the dipole-dipole interaction takes the form
\begin{equation} \label{17}
\begin{array}{l}
\displaystyle{%
  \hbar(i-\gamma)\dot\psi= -\frac{\hbar^2}{2m}\triangle\psi+\big(d_sE_0-{\bf E}\!\cdot\!{\bf d}\big)\psi+g\big(|\psi|^2-n_0\big)\psi + \psi\int\! U_D({\bf r},{\bf r}')\big|\psi({\bf r}',t)\big|^2d{\bf r}'. %
}
\end{array}
\end{equation}

The polarization vector
\begin{equation} \label{18}
\begin{array}{l}
\displaystyle{%
   {\bf P}={\bf d}\,|\psi|^2
}
\end{array}
\end{equation}
is also a sum of constant and alternating terms ${\bf P}={\bf
P}_s+\delta{\bf P}(t)$. The constant polarization is a sum of the
spontaneous polarization of the condensate of atoms with intrinsic
dipole moments and the polarization under the action of the constant
external field:
\begin{equation} \label{19}
\begin{array}{l}
\displaystyle{%
   {\bf P}_s=d_0n_0{\bf e}+\alpha_0n_0E_0{\bf e}=d_s n_0{\bf e},
}
\end{array}
\end{equation}
and the nonstationary contribution into the total vector of
polarization has the form
\begin{equation} \label{20}
\begin{array}{l}
\displaystyle{%
   \delta{\bf P}=\psi_0d_s\big(\delta\psi+\delta\psi^*\big){\bf e} + n_0 \int_{0}^\infty\!\alpha(\tau)\delta{\bf E}(t-\tau)d\tau. %
}
\end{array}
\end{equation}
Thus, the total electric displacement ${\bf D}={\bf E}+4\pi{\bf P}$
is a sum of the constant displacement ${\bf D}_s=4\pi{\bf P}_s+{\bf
E}_0=\varepsilon_0{\bf E}_0+4\pi d_0n_0{\bf e}$\, and the
displacement induced by the alternating field
\begin{equation} \label{21}
\begin{array}{l}
\displaystyle{%
   {\bf D}={\bf D}_s+\delta{\bf D},
}
\end{array}
\end{equation}
where $\delta{\bf D}=\delta{\bf E}+4\pi\delta{\bf P}$ is the
contribution into the displacement that depends on time,
$\varepsilon_0=1+4\pi\alpha_0n_0$ is the static permittivity.

\section{Permittivity of BEC atoms with a dipole moment }\vspace{-0mm}
The relation (\ref{20}) and equation (\ref{17}) allow to obtain the
permittivity of BEC atoms with a dipole moment as a function of the
frequency and the wave vector. It should be noted that the de
Broglie wavelength of the condensate atoms at low temperatures is
large and greatly exceeds the mean interatomic distance, so that all
atoms are collectivized and the many-particle system actually loses
its discrete structure. For this reason BEC can be considered in a
good approximation as a continuous medium, described by the
macroscopic wave function even at high frequencies, when the
wavelength of propagating oscillations is of the order or less than
the mean distance between particles. Therefore we will use the
equation (\ref{17}) also for the description of the short sound
waves at high frequencies comparable with the internal frequencies
of an atom.

Let us linearize the equation (\ref{17}), taking into account the
terms linear in the alternating field and the fluctuation of the
dipole moment and assuming $\delta\psi=\psi-\psi_0$. Writing the
dipole moment of an atom in the form ${\bf d}({\bf r})=d_s{\bf
e}+\delta{\bf d}({\bf r})$, we find accurate to the linear terms in
the fluctuation of the dipole moment the expression for the
dipole-dipole potential $U_D({\bf r},{\bf r}')\approx
U_D^{(0)}+U_D^{(1)}({\bf r},{\bf r}')$, where the contributions of
the zero and first approximations have the form
\begin{equation} \label{22}
\begin{array}{c}
\displaystyle{%
  U_D^{(0)}=\frac{d_s^2}{R^3}\Big[1-3\big({\bf x}\cdot{\bf e}\big)^2\Big], %
}\vspace{2mm}\\ %
\displaystyle{\hspace{0mm}%
  U_D^{(1)}=\frac{d_s^2}{R^3}\big[{\bf e}\cdot\delta{\bf d}({\bf r})+{\bf e}\cdot\delta{\bf d}({\bf r}')\big] %
  -3\frac{d_s^2}{R^3}\big({\bf e}\cdot{\bf x}\big)\big[{\bf x}\cdot\delta{\bf d}({\bf r})+{\bf x}\cdot\delta{\bf d}({\bf r}')\big]. %
}
\end{array}
\end{equation}
In the linearized equation it is convenient to pass to the real
quantities
\begin{equation} \label{23}
\begin{array}{l}
\displaystyle{%
   \delta\Psi\equiv\delta\psi+\delta\psi^*, \qquad \delta\Phi\equiv i\big(\delta\psi-\delta\psi^*\big).%
}
\end{array}
\end{equation}
If in the macroscopic wave function one singles out the modulus and
the phase $\psi=\rho e^{i\chi}$, assuming in the equilibrium
$\rho_0=\psi_0$ and $\chi_0=0$, then it turns out that the functions
introduced in (\ref{23}) are connected with the fluctuations of the
modulus $\delta\rho$ and the phase $\delta\chi$ by the relations
$\delta\Psi=2\delta\rho$, $\delta\Phi=2\psi_0\delta\chi$. The
linearized system of equations for the real functions (\ref{23})
takes the form
\begin{equation} \label{24}
\begin{array}{c}
\displaystyle{%
   \hbar\delta\dot{\Phi}-\hbar\gamma\delta\dot{\Psi}=-\frac{\hbar^2}{2m}\triangle\delta\Psi+2gn_0\delta\Psi
   -2\psi_0\big(d_s{\bf e}\!\cdot\!\delta{\bf E}+E_0{\bf e}\!\cdot\!\delta{\bf d}\big)+2\delta J_D,  %
}\vspace{2mm}\\ %
\displaystyle{\hspace{0mm}%
   \hbar\delta\dot{\Psi}+\hbar\gamma\delta\dot{\Phi}=\frac{\hbar^2}{2m}\triangle\delta\Phi,   %
}
\end{array}
\end{equation}
where
\begin{equation} \label{25}
\begin{array}{l}
\displaystyle{%
   \delta J_D=n_0\psi_0d_s\Bigg\{\int\frac{\big[{\bf e}\!\cdot\!\delta{\bf d}({\bf r})+{\bf e}\!\cdot\!\delta{\bf d}({\bf r}')\big]}{R^3}d{\bf r}' %
   -3\int\frac{\big({\bf e}\cdot{\bf x}\big)\big[{\bf x}\!\cdot\!\delta{\bf d}({\bf r})+{\bf x}\!\cdot\!\delta{\bf d}({\bf r}')\big]}{R^3}d{\bf r}'  \Bigg\}\,+  %
}\vspace{2mm}\\ %
\displaystyle{\hspace{7mm}%
  +\, n_0d_s^2\int\frac{\Big[1-3\big({\bf e}\cdot{\bf x}\big)^2\Big]}{R^3}\delta\Psi({\bf r}')d{\bf r}'.   %
}
\end{array}
\end{equation}
This system of equations describes in the linear approximation the
dynamics of the superfluid system in the weak electric field. Let us
assume that the alternating electric field varies according to the
harmonic law $\delta{\bf E}({\bf r},t)={\bf b}e^{iQ({\bf r},t)}$,
where $Q({\bf r},t)\equiv{\bf k}{\bf r}-\omega t$, ${\bf k}$ is the
wave vector, $\omega$ is the frequency, ${\bf b}$ is the constant
complex vector. The solutions of the system of equations (\ref{24})
also have the form $\delta\Psi({\bf r},t)=\Psi_0e^{iQ({\bf r},t)}$,
$\delta\Phi({\bf r},t)=\Phi_0e^{iQ({\bf r},t)}$. Since the
fluctuations of the oscillating quantities are chosen in the form of
plane waves, then $\delta{\bf d}({\bf r}')=\delta{\bf d}({\bf
r})e^{-i{\bf k}({\bf r}-{\bf r}')}$, $\delta\Psi({\bf
r}')=\delta\Psi({\bf r})e^{-i{\bf k}({\bf r}-{\bf r}')}$. In view of
this we find
\begin{equation} \label{26}
\begin{array}{l}
\displaystyle{%
   \delta J_D=-\frac{4\pi d_s n_0}{3}\big(\delta_{ij}-3s_is_j\big)  %
   \big[\psi_0e_i\delta d_j({\bf r})+d_s e_i e_j \delta\Psi({\bf r})\big].
}
\end{array}
\end{equation}
As a result, we come to the system of linear equations
\begin{equation} \label{27}
\begin{array}{c}
\displaystyle{%
   -i\hbar\omega\Phi_0+\big(i\hbar\omega\gamma-\xi_k-2gn_0+u \big)\Psi_0=-2\psi_0\big[d_s+\alpha(\omega)E_0\big]{\bf e}\cdot{\bf b}+{\bf L}\cdot{\bf b},   %
}\vspace{2mm}\\ %
\displaystyle{\hspace{0mm}%
  \big(i\hbar\omega\gamma-\xi_k \big)\Phi_0+i\hbar\omega\Psi_0=0,  %
}
\end{array}
\end{equation}
where the quantities
\begin{equation} \label{28}
\begin{array}{c}
\displaystyle{%
   u\equiv\frac{8\pi d_s^2n_0}{3}\Big[1-3\big({\bf e}\cdot{\bf s}\big)^2\Big],   %
}\vspace{2mm}\\ %
\displaystyle{\hspace{0mm}%
  {\bf L}\equiv -2\psi_0\!\left[\frac{4\pi d_sn_0}{3}\alpha(\omega)\right]\!{\bf e} +8\pi\psi_0n_0d_s\alpha(\omega)\big({\bf e}\cdot{\bf s}\big){\bf s} %
}
\end{array}
\end{equation}
appeared due to taking into account of the dipole-dipole
interaction, and à $\xi_k=\hbar^2k^2\big/2m$. The solutions of the
system of equations (\ref{27}) have the form
\begin{equation} \label{29}
\begin{array}{c}
\displaystyle{%
   \Psi_0=\frac{ \big(\xi_k-i\gamma\hbar\omega\big)}{D}\big\{\!-\!2\psi_0\big[d_s+\alpha(\omega)E_0\big]{\bf b}\cdot{\bf e}+{\bf b}\cdot{\bf L}\big\},   %
}\vspace{2mm}\\ %
\displaystyle{\hspace{0mm}%
  \Phi_0=\frac{ i\hbar\omega}{D}\big\{\!-\!2\psi_0\big[d_s+\alpha(\omega)E_0\big]{\bf b}\cdot{\bf e}+{\bf b}\cdot{\bf L}\big\},  %
}
\end{array}
\end{equation}
and here
\begin{equation} \label{30}
\begin{array}{c}
\displaystyle{%
   D\equiv D(\omega,{\bf k})\equiv (\hbar\omega)^2- \big(i\hbar\omega\gamma-\xi_k\big)\big(i\hbar\omega\gamma-\xi_k-2gn_0+u\big).  %
}
\end{array}
\end{equation}
Taking into account the relations (\ref{29}), and that the amplitude
of oscillations of the polarization vector (\ref{20}) ${\bf
p}=\psi_0d_s\Psi_0{\bf e}+n_0\alpha(\omega){\bf b}$, we find the
form of the polarization tensor of BEC with allowance for the
dipole-dipole interaction
\begin{equation} \label{31}
\begin{array}{c}
\displaystyle{%
   \kappa_{ij}(\omega,{\bf k})=\kappa_0(\omega)\delta_{ij}+\kappa_\perp(\omega,k)e_ie_j+\kappa_d(\omega,k)({\bf e}\cdot{\bf s})e_is_j,  %
}
\end{array}
\end{equation}
where
\begin{equation} \label{32}
\begin{array}{c}
\displaystyle{%
   \kappa_0(\omega)=n_0\alpha(\omega),   %
}\vspace{2mm}\\ %
\displaystyle{\hspace{0mm}%
  \kappa_\perp(\omega,k)=-2n_0d_s\big(\xi_k-i\hbar\omega\gamma\big)D^{-1}\left[d_s+\alpha(\omega)E_0+\frac{4\pi d_sn_0}{3}\alpha(\omega)\right],   %
}
\vspace{2mm}\\ %
\displaystyle{\hspace{0mm}%
  \kappa_d(\omega,k)= 8\pi d_s^2n_0^2\alpha(\omega)(\xi_k-i\hbar\omega\gamma\big)D^{-1}. %
}
\end{array}
\end{equation}
In neglect of the dipole-dipole interaction in the formula
(\ref{31}) we have to put
$\kappa_\perp(\omega,k)=-2n_0d_s\big(\xi_k-i\hbar\omega\gamma\big)D^{-1}\left[d_s+\alpha(\omega)E_0\right]$
and $\kappa_d(\omega,k)=0$. Thus, in this case the tensor of
permittivity $\varepsilon_{ij}(\omega,{\bf
k})=\delta_{ij}+4\pi\kappa_{ij}(\omega,{\bf k})$ can be presented in
the form
\begin{equation} \label{33}
\begin{array}{c}
\displaystyle{%
   \varepsilon_{ij}(\omega,{\bf k})=\varepsilon_{ij}^s(\omega,{\bf k})+\varepsilon_{ij}^a(\omega,{\bf k}),  %
}
\end{array}
\end{equation}
where the symmetric and antisymmetric parts of the tensor are
singled out:
\begin{equation} \label{34}
\begin{array}{c}
\displaystyle{%
   \varepsilon_{ij}^s(\omega,{\bf k})=\varepsilon_{ji}^s(\omega,{\bf k}) =   %
   \big[1+4\pi\kappa_0(\omega)\big]\delta_{ij}+4\pi\kappa_\perp(\omega,k)e_ie_j+2\pi\kappa_d(\omega,k)({\bf e}\cdot{\bf s})\big(e_is_j+e_js_i\big), %
}\vspace{2mm}\\ %
\displaystyle{\hspace{0mm}%
  \varepsilon_{ij}^a(\omega,{\bf k})=-\varepsilon_{ji}^a(\omega,{\bf k})=2\pi\kappa_d(\omega,k)({\bf e}\cdot{\bf s})\big(e_is_j-e_js_i\big).   %
}
\end{array}
\end{equation}
As it is seen, the tensor of permittivity is a function of both the
frequency and the wave vector, so that the medium possesses both the
frequency and the spatial dispersion. The account of the
dipole-dipole interaction results in that the tensor of permittivity
ceases to be symmetric. Note that in the most general case the
tensor of permittivity is not required to be either hermitian or
symmetric [25]. The proof of its symmetry, based on the principle of
symmetry of the kinetic coefficients [26], is not applicable in this
case. In the presence of the antisymmetric additional term to the
hermitian tensor of permittivity, for example in the magnetic field,
the medium becomes optically active or gyrotropic [26]. The phase
velocities of waves with the right and left circular polarization
differ in the gyrotropic medium, which leads to the rotation of the
plane of polarization of the linearly polarized wave. In our case in
the absence of dissipation the tensor of permittivity proves to be
real and unsymmetric. However, as will be seen, this does not lead
to the effects of rotation of the plane of polarization.

A Fourier component of the polarizability of an atom
$\alpha(\omega)\equiv\int_0^\infty\!\alpha(\tau)e^{i\omega\tau}d\tau$
is often chosen in the model of the damped oscillator in the form
\begin{equation} \label{35}
\begin{array}{c}
\displaystyle{%
   \alpha(\omega)=\frac{N_ae^2}{m_0}\frac{1}{\omega_0^2-\omega^2-i\nu\omega},  %
}
\end{array}
\end{equation}
where $N_a$ is a number of electrons in an atom which give a
contribution to the polarization, $\nu$ is a phenomenological
coefficient of damping, $m_0$ is the mass of an electron, $\omega_0$
is the characteristic resonance frequency of oscillations of an
electron in an atom. In this model the static polarizability of an
atom $\alpha_0\equiv\alpha(0)=N_ae^2\big/m_0\omega_0^2$. The real
and imaginary parts of the polarizability (\ref{35})
$\alpha(\omega)=\alpha'(\omega)+i\alpha''(\omega)$:
\begin{equation} \label{36}
\begin{array}{c}
\displaystyle{%
   \alpha'(\omega)=\alpha_0\frac{\omega_0^2\big(\omega_0^2-\omega^2\big)}{\big(\omega_0^2-\omega^2\big)^2+\nu^2\omega^2}, \qquad%
   \alpha''(\omega)=\alpha_0\frac{\nu\omega\omega_0^2}{\big(\omega_0^2-\omega^2\big)^2+\nu^2\omega^2}. %
}
\end{array}
\end{equation}

Far from the resonant frequency $\omega_0$ the mutual influence of
the sound waves in the condensate and the electromagnetic waves
turns out to be small, so we pay the main attention to the
propagation of waves near the resonant frequency. We are going to
consider the propagation of electromagnetic waves of a certain
frequency, assuming $\omega$ to be real. Owing to the presence of
damping the wave vector of the propagating wave is complex ${\bf
k}={\bf k}'+i{\bf k}''$. Generally speaking, the real and imaginary
parts of this vector can be directed differently. In the following
we confine ourselves to considering the uniform wave, for which
${\bf k}=(k'+ik''){\bf s}$, where ${\bf s}$ is the unit vector
determining the direction of propagation of the wave.

\section{Dispersion equation for waves in BEC}\vspace{-0mm}
In neglect of the magnetic properties of the medium, from the
Maxwell system of equations
\begin{equation} \label{37}
\begin{array}{c}
\displaystyle{%
   {\rm div}\, \delta{\bf D}=0, \qquad {\rm div}\, \delta{\bf B}=0, %
}\vspace{2mm}\\ %
\displaystyle{\hspace{0mm}%
  {\rm rot}\, \delta{\bf B}=\frac{1}{c}\frac{\partial \delta{\bf D}}{\partial t}, \qquad %
  {\rm rot}\, \delta{\bf E}=-\frac{1}{c}\frac{\partial \delta{\bf B}}{\partial t},  %
}
\end{array}
\end{equation}
there follows the system of linear equations describing the
propagation of the electromagnetic waves in BEC
\begin{equation} \label{38}
\begin{array}{c}
\displaystyle{%
   \left[k^2 \big(\delta_{ij}-s_is_j\big)-\frac{\omega^2}{c^2}\varepsilon_{ij}\right]\!b_j=0.  %
}
\end{array}
\end{equation}
For the dipole moment oriented along the vector ${\bf e}=(0,0,1)$
and the direction of propagation of the wave ${\bf
s}=(\sin\theta,0,\cos\theta)$, the tensor of permittivity has the
following components
\begin{equation} \label{39}
\begin{array}{c}
\displaystyle{%
   \varepsilon_{xx}=\varepsilon_{yy}=\varepsilon, \qquad  \varepsilon_{zz}=\varepsilon+ 4\pi\big(\kappa_\perp+ \kappa_d\cos^2\theta\big), %
}\vspace{2mm}\\ %
\displaystyle{\hspace{0mm}%
  \varepsilon_{xy}=\varepsilon_{yx}=\varepsilon_{yz}=\varepsilon_{zy}=0,  %
}\vspace{2mm}\\ %
\displaystyle{\hspace{0mm}%
  \varepsilon_{zx}=2\pi\kappa_d\sin\theta\cos\theta, \qquad \varepsilon_{xz}=0.  %
}
\end{array}
\end{equation}
Here $\varepsilon\equiv\varepsilon(\omega)=1+4\pi\kappa_0(\omega),
\kappa_\perp\equiv\kappa_\perp(\omega,k),
\kappa_d\equiv\kappa_d(\omega,k)$. As we see, at $\theta=0$ and
$\theta=\pi/2$ the antisymmetric component of the tensor of
permittivity becomes zero.

It proves more convenient to pass from the oscillations of the
electric field intensity to the oscillations of the electric
displacement $\delta{\bf D}({\bf r},t)={\bf f}e^{iQ({\bf r},t)}$, so
that $f_i=\varepsilon_{ij}b_j$. The reciprocal transformation is
$b_i=\eta_{ij}f_j$, where $\eta_{ij}\equiv\varepsilon_{ij}^{-1}$ is
the inverse tensor of permittivity:
$\eta_{ik}\varepsilon_{kj}=\delta_{ij}$ . This tensor has the form
\begin{equation} \label{40}
\begin{array}{c}
\displaystyle{%
   \eta_{ij}\equiv\eta_{ij}(\omega,{\bf k})=\frac{1}{\varepsilon}
   \left[\delta_{ij}-4\pi\frac{\kappa_\perp e_ie_j+\kappa_d({\bf e}\cdot{\bf s})e_is_j}
   {\varepsilon+4\pi\kappa_\perp+4\pi\kappa_d({\bf e}\cdot{\bf s})^2} \right].  %
}
\end{array}
\end{equation}
The inverse tensor (\ref{40}) is also unsymmetric and has the
following components:
\begin{equation} \label{41}
\begin{array}{c}
\displaystyle{%
   \eta_{xx}=\eta_{yy}=\frac{1}{\varepsilon}, \qquad  \eta_{zz}=\frac{1}{\varepsilon+4\pi\kappa_\perp+4\pi\kappa_d\cos^2\theta}, \qquad %
   \eta_{xy}=\eta_{yx}=\eta_{yz}=\eta_{zy}=0,
}\vspace{2mm}\\ %
\displaystyle{\hspace{0mm}%
  \eta_{xz}=0,\qquad \eta_{zx}= -\frac{4\pi\kappa_d\sin\theta\cos\theta}
   {\varepsilon\big(\varepsilon+4\pi\kappa_\perp+4\pi\kappa_d\cos^2\theta\big)}. %
}
\end{array}
\end{equation}
The amplitudes of oscillations of the displacement satisfy the
system of linear equations following from (\ref{38})
\begin{equation} \label{42}
\begin{array}{c}
\displaystyle{%
      \left(\frac{\omega^2}{c^2}\delta_{ik}-k^2\vartheta_{ik}\right)\!f_k=0,  %
}
\end{array}
\end{equation}
where $\vartheta_{ik}\equiv\big(\delta_{ij}-s_is_j\big)\eta_{jk}$.
Since $s_i\vartheta_{ik}=0$, then from (\ref{42}) it is obvious that
${\bf s}\cdot{\bf f}=0$ and therefore the vector of the displacement
lies in the plane perpendicular the direction of propagation of the
wave. The components of the tensor $\vartheta_{ik}$ have the form
\begin{equation} \label{43}
\begin{array}{c}
\displaystyle{%
   \vartheta_{xx}=\frac{\cos^2\theta\big(\varepsilon+4\pi\kappa_\perp+4\pi\kappa_d\big)}{\varepsilon\big(\varepsilon+4\pi\kappa_\perp+4\pi\kappa_d\cos^2\theta\big)}, \qquad  %
   \vartheta_{yy}=\frac{1}{\varepsilon}, \qquad
   \vartheta_{zz}=\frac{\sin^2\theta}{\big(\varepsilon+4\pi\kappa_\perp+4\pi\kappa_d\cos^2\theta\big)}, %
}\vspace{2mm}\\ %
\displaystyle{\hspace{0mm}%
  \vartheta_{xy}=\vartheta_{yx}=\vartheta_{yz}=\vartheta_{zy}=0, %
}\vspace{2mm}\\ %
\displaystyle{\hspace{0mm}%
  \vartheta_{xz}=-\frac{\sin\theta\cos\theta}{\big(\varepsilon+4\pi\kappa_\perp+4\pi\kappa_d\cos^2\theta\big)}, \qquad %
  \vartheta_{zx}=-\frac{\sin\theta\cos\theta\big(\varepsilon+4\pi\kappa_\perp+4\pi\kappa_d\big)}{\varepsilon\big(\varepsilon+4\pi\kappa_\perp+4\pi\kappa_d\cos^2\theta\big)}.%
}
\end{array}
\end{equation}
From the condition
\begin{equation} \label{44}
\begin{array}{c}
\displaystyle{%
      {\rm det}\!\left[k^2\vartheta_{ik}-\frac{\omega^2}{c^2}\delta_{ik}\right]=0  %
}
\end{array}
\end{equation}
we find the equation determining the dispersion laws for the
propagating waves:
\begin{equation} \label{45}
\begin{array}{c}
\displaystyle{%
      \left(k^2\vartheta_{yy}-\frac{\omega^2}{c^2}\right)\!  %
      \left[ \left(k^2\vartheta_{xx}-\frac{\omega^2}{c^2}\right)\!\left(k^2\vartheta_{zz}-\frac{\omega^2}{c^2}\right)\!-k^4\vartheta_{xz}\vartheta_{zx}
      \right]=0.
}
\end{array}
\end{equation}
It follows from here that the wave, in which oscillates the
component $f_y$ perpendicular to the plane containing the vectors
${\bf e}$ and ${\bf s}$, propagates with the dispersion law
\begin{equation} \label{46}
\begin{array}{c}
\displaystyle{%
     \frac{\omega^2}{c^2}\,\varepsilon=k^2,  %
}
\end{array}
\end{equation}
and oscillations of the condensate are absent in this wave. For the
wave in which the displacement vector oscillates in the plane
$(x,z)$, where the vectors ${\bf e}$ and ${\bf s}$ lie, the
dispersion law, according to (\ref{45}), is determined by the
equation
\begin{equation} \label{47}
\begin{array}{c}
\displaystyle{%
      \left(\frac{\omega^2}{c^2k^2}-\vartheta_{xx}\right)\!\left(\frac{\omega^2}{c^2k^2}-\vartheta_{zz}\right)\!-\vartheta_{xz}\vartheta_{zx}=0.  %
}
\end{array}
\end{equation}
Since, as can be seen from formulas (\ref{43}),
$\vartheta_{xx}\vartheta_{zz}=\vartheta_{xz}\vartheta_{zx}$, then
from (\ref{47}) apart from the trivial solution $\omega^2=0$ there
follows the equation determining the dispersion law
\begin{equation} \label{48}
\begin{array}{c}
\displaystyle{%
      \frac{\omega^2}{c^2k^2}\,\varepsilon=1-\frac{4\pi\kappa_\perp\sin^2\theta}{\varepsilon+4\pi\kappa_\perp+4\pi\kappa_d\cos^2\theta}.  %
}
\end{array}
\end{equation}
As seen, the nontrivial effects under propagation of waves arise if
the wave propagates at the angle to the direction of the dipole
moments, so in the following we consider the cases when
$\vartheta\neq 0,\pi$. The relation between the components of the
displacement vector in the plane $(x,z)$ is determined by the system
of linear equations
\begin{equation} \label{49}
\begin{array}{c}
\displaystyle{%
      \left(\frac{\omega^2}{c^2}-k^2\vartheta_{xx}\right)\!f_x-k^2\vartheta_{xz}f_z=0,  %
}\vspace{2mm}\\ %
\displaystyle{\hspace{0mm}%
      k^2\vartheta_{zx}f_x-\left(\frac{\omega^2}{c^2}-k^2\vartheta_{zz}\right)\!f_z=0. %
}
\end{array}
\end{equation}
Taking into account formulas (\ref{32}), the equation (\ref{48}),
which determines the dispersion laws for the sound waves in BEC and
the electromagnetic waves with account of their damping, can be
written in the form
\begin{equation} \label{50}
\begin{array}{c}
\displaystyle{%
      \left[k^2-\frac{\omega^2}{c^2}\varepsilon(\omega)\right]\!\Big\{\!(\hbar\omega)^2-\big(i\hbar\omega\gamma-\xi_k\big)\big(i\hbar\omega\gamma-\xi_k-2gn_0+u\big)\!\Big\}=  %
}\vspace{2mm}\\ %
\displaystyle{\hspace{0mm}%
      =\big(i\hbar\omega\gamma-\xi_k\big)\Bigg\{8\pi n_0d_s\big[d_s+\alpha(\omega)E_0\big]\!\left(\frac{\omega^2}{c^2}-\cos^2\theta\frac{k^2}{\varepsilon(\omega)}\right)+ %
}\vspace{2mm}\\ %
\displaystyle{\hspace{0mm}%
   +\,32\pi^2d_s^2n_0^2\alpha(\omega)\!\left[\frac{\omega^2}{c^2}\!\left(\frac{1}{3}-\cos^2\theta\right)\!+\frac{2}{3}\cos^2\theta\frac{k^2}{\varepsilon(\omega)}\right]\!\!\Bigg\} %
}
\end{array}
\end{equation}
where, as in (\ref{28}),  it is designated $u\equiv 8\pi
d_s^2n_0\big(1/3-\cos^2\theta\big)$. In the following we consider
the external electric field weak, assuming $E_0\rightarrow 0$, so
that $d_s=d_0$. Actually, in the problem under consideration the
external field is needed only to single out a direction along which
the dipoles are oriented. The account for the influence of the
external static field on the propagation of waves in the condensate
is also of interest and needs a separate research. Neglecting the
quantity $u$ in the left side of equation (\ref{50}) and the second
term in the curly brackets in the right side of this equation, we
obtain the equation which determines the dispersion laws of waves
without account of the dipole-dipole interaction.

It is convenient to represent the obtained dispersion equation in
the dimensionless form. It is natural to define the two
characteristic wave numbers and the dimensionless frequency:
\begin{equation} \label{51}
\begin{array}{c}
\displaystyle{%
    k_1\equiv\frac{\omega_0}{c},\qquad k_2\equiv \sqrt{\frac{2m\omega_0}{\hbar}},\qquad x\equiv\frac{\omega}{\omega_0}.  %
}
\end{array}
\end{equation}
Then in the zero constant electric field $E_0=0$ the equation
(\ref{50}) takes the form
\begin{equation} \label{52}
\begin{array}{c}
\displaystyle{%
      \left[\left(\frac{k}{k_1}\right)^{\!2}-x^2\varepsilon'(x)-ix^2\varepsilon''(x)\right]\!\!  %
      \left[\left(\frac{k}{k_2}\right)^{\!2}-\beta_-(x)-i\gamma x\right]\!\!  %
      \left[\left(\frac{k}{k_2}\right)^{\!2}+\beta_+(x)-i\gamma x\right] =-\eta_B\Omega\big(k^2,x\big), %
}
\end{array}
\end{equation}
where
\begin{equation} \label{53}
\begin{array}{c}
\displaystyle{%
      \Omega\big(k^2,x\big)\equiv\left[ix\gamma-\left(\frac{k}{k_2}\right)^{\!2}\right]\!\!  %
      \Bigg\{x^2\Big[1+\big(\varepsilon(x)-1\big)f_D(\theta)\Big]+
      \frac{5}{3}\cos^2\theta\left(\frac{k}{k_1}\right)^{\!2}\!\left[\frac{2}{5}-\frac{1}{\varepsilon(x)}\right]
      \!\!\Bigg\}.
}
\end{array}
\end{equation}
Here the designation for the function $f_D(\theta)\equiv
1/3-\cos^2\theta$ is introduced, being typical for the systems with
the dipole-dipole interaction, and also the designation for the
function of the dimensionless frequency
\begin{equation} \label{54}
\begin{array}{c}
\displaystyle{%
    \beta_\pm(x)\equiv\sqrt{\frac{\phi^2}{4}+x^2}\pm \frac{\phi}{2},  %
}
\end{array}
\end{equation}
where $\phi\equiv\phi(\theta)\equiv 2\upsilon-\eta_Bf_D(\theta)$,
the dimensionless parameter $\upsilon\equiv gn_0\big/\hbar\omega_0$
is determined by the constant $g$ of the short-range interaction of
atoms of the condensate. Also (\ref{52}) contains the dimensionless
constant which characterizes the strength of the interaction of the
condensate of particles with a dipole moment with the electric
field:
\begin{equation} \label{55}
\begin{array}{c}
\displaystyle{%
    \eta_B\equiv\frac{8\pi d_0^2n_0}{\hbar\omega_0}.  %
}
\end{array}
\end{equation}
In (\ref{52}) the real and imaginary parts of the permittivity,
depending on the dimensionless frequency, are given by the formulas
\begin{equation} \label{56}
\begin{array}{c}
\displaystyle{%
   \varepsilon'(x)=1+\big(\varepsilon_0-1\big)\frac{\big(1-x^2\big)}{\big(1-x^2\big)^2+\tilde{\nu}^2x^2}, \qquad%
   \varepsilon''(x)=\big(\varepsilon_0-1\big)\frac{\tilde{\nu}x}{\big(1-x^2\big)^2+\tilde{\nu}^2x^2}, %
}
\end{array}
\end{equation}
where $\tilde{\nu}\equiv \nu\big/\omega_0$. In the presence of
damping the wave number is complex $k=k'+i k''$.

Let us give numerical estimates of the introduced quantities.
Assuming $m\approx 10^{-23}\,g$, $\omega_0\approx 10^{15}\,Hz$, we
have $k_1\sim 10^4\div 10^5\,cm^{-1}$ and $k_2\sim 10^9\div
10^{10}\,cm^{-1}$, so that $k_2/k_1\sim 10^5$. Since the Bogolyubov
speed of sound in the condensate $c_B$ is determined by the formula
$c_B^2=gn_0/m$, then $\upsilon\equiv mc_B^2\big/\hbar\omega_0$. In
the atomic condensates $c_B\approx 1\,sm/s$ [27,28], so that
$\upsilon\sim 10^{-11}$. If we take $c_B\sim 10^4\,sm/s$ -- the
order of magnitude of the first speed of sound in helium, then
$\upsilon\sim 10^{-3}$. Let us give an estimate for the value of the
parameter (\ref{55}). The dipole moment of an atom can be written in
the form $d_0=ad_D$, where $d_D=1D=10^{-18}\,cm^{5/2}g^{1/2}s^{-1}$
is a typical value of the dipole moment for the polar molecules, $a$
is a dimensionless constant. In the atomic condensates the
characteristic density $n_0\approx 10^{14}\,cm^{-3}$, so that
$\eta_B\sim 10^{-9}a^2$. In this case this parameter is small, even
if the dipole moment of an atom is of the order of the dipole moment
of the polar molecule, when $a\sim 1$. For the particle number
density close to the density in the liquid helium $n_0\approx
10^{22}\,cm^{-3}$ we obtain $\eta_B\sim 10^{-1}a^2$. In works
[12,13] there were given arguments in favor of that at some
temperature somewhat higher than the temperature of the superfluid
transition, the liquid helium transforms into the polarized state.
In this case atoms of the helium in the medium acquire the intrinsic
dipole moment. A rough estimate in [12,13] gave the value of this
moment which proved to be by 5\,--\,6 orders of magnitude less than
the characteristic dipole moment of the polar molecules, so that the
parameter $a\sim 10^{-6}\div 10^{-5}$, which gives $\eta_B\sim
10^{-13}\div 10^{-11}$. Thus, we can always consider $\eta_B\ll 1$.

\section{Solution of dispersion equation for waves in BEC }\vspace{-0mm}
In this section we obtain the solution of the dispersion equation
(\ref{52}), taking into account a small value of the parameter
$\eta_B$. In view of the fact that in general case the wave number
is complex $k=k'+i k''$, let us introduce the following notations:
\begin{equation} \label{57}
\begin{array}{c}
\displaystyle{%
    z^2\equiv  \frac{\left(k'^2-k''^2\right)}{k_1k_2},\qquad y^2\equiv\frac{k'k''}{k_1k_2},\qquad \kappa\equiv\frac{k_1}{k_2},   %
}\vspace{2mm}\\ %
\displaystyle{\hspace{0mm}%
     \left(\frac{k}{k_1}\right)^{\!2}=\frac{1}{\kappa}\left(z^2+2iy^2\right),\qquad %
     \left(\frac{k}{k_2}\right)^{\!2}=\kappa\left(z^2+2iy^2\right). %
}
\end{array}
\end{equation}
The real and imaginary parts of the wave number are expressed
through the parameters $z$ and $y$ by the relations:
\begin{equation} \label{58}
\begin{array}{c}
\displaystyle{%
    k'^2\big/k_1k_2=\frac{1}{2}\Big[\sqrt{z^4+4y^4}+z^2\Big], \qquad  %
    k''^2\big/k_1k_2=\frac{1}{2}\Big[\sqrt{z^4+4y^4}-z^2\Big].  %
}
\end{array}
\end{equation}
In terms of variables (\ref{57}) the original equation (\ref{52}) takes the form %
\begin{equation} \label{59}
\begin{array}{c}
\displaystyle{%
      \left[\left(\frac{z^2}{\kappa}-x^2\varepsilon'(x)\!\right)+i\left(2\frac{y^2}{\kappa}-x^2\varepsilon''(x)\!\right)\right]\!\!  %
      \Big[\!\left(\kappa z^2-\beta_-(x)\right)+i\!\left(2\kappa y^2-\gamma x\right)\!\Big]\times  %
}\vspace{2mm}\\ %
\displaystyle{\hspace{0mm}%
     \times \Big[\!\left(\kappa z^2+\beta_+(x)\right)+i\!\left(2\kappa y^2-\gamma x\right)\!\Big] %
       =-\eta_B\Omega\big(z^2,y^2,x\big), %
}
\end{array}
\end{equation}
where now
\begin{equation} \label{60}
\begin{array}{c}
\displaystyle{%
      \Omega\big(z^2,y^2,x\big)\equiv\Big[ix\gamma-\kappa\left(z^2+2iy^2\right)\!\Big] %
      \Bigg\{x^2\Big[1+\big(\varepsilon(x)-1\big)f_D(\theta)\Big]+
      \frac{5}{3\kappa}\cos^2\theta\left(z^2+2iy^2\right)\!\left[\frac{2}{5}-\frac{1}{\varepsilon(x)}\right]\!\!\Bigg\}.
}
\end{array}
\end{equation}
Since the damping is assumed to be sufficiently weak and the
parameter $\eta_B$ is small, then the dissipative terms in function
(\ref{60}) are neglected, so that it takes the form
\begin{equation} \label{61}
\begin{array}{c}
\displaystyle{%
      \Omega\big(z^2,x\big)=-\kappa z^2 %
      \Bigg\{x^2\Big[1+\big(\varepsilon'(x)-1\big)f_D(\theta)\Big]+
      \frac{5}{3\kappa}z^2\cos^2\theta\left[\frac{2}{5}-\frac{1}{\varepsilon'(x)}\right]\!\!\Bigg\}.
}
\end{array}
\end{equation}
In the left part of equation (\ref{59}) the terms quadratic in the
dissipative coefficients in the real part are neglected, and the
imaginary part, due to the fact that the right part is real, is
equated to zero. As a result we come to the equation
\begin{equation} \label{62}
\begin{array}{c}
\displaystyle{%
      \left(\frac{z^2}{\kappa}-x^2\varepsilon'(x)\!\right)\!\left(\kappa z^2-\beta_-(x)\right)\!\left(\kappa z^2+\beta_+(x)\right)
      =-\eta_B\Omega\big(z^2,x\big), %
}
\end{array}
\end{equation}
and the function describing the damping of waves is expressed
through the solutions of equation (\ref{62}):
\begin{equation} \label{63}
\begin{array}{c}
\displaystyle{%
      y^2(x)=\frac{\kappa x^2\varepsilon''(x)\!\left(\kappa z^2-\beta_-(x)\right)\!\left(\kappa z^2+\beta_+(x)\right)
      +\gamma x\!\left(z^2-\kappa x^2\varepsilon'(x)\right)\!\left(2\kappa z^2+\phi\right)}
      {2\big[\!\left(\kappa z^2-\beta_-(x)\right)\!\left(\kappa z^2+\beta_+(x)\right)
      +\kappa\!\left(z^2-\kappa x^2\varepsilon'(x)\right)\!\left(2\kappa z^2+\phi\right)\!\big]}. %
}
\end{array}
\end{equation}
In neglect of damping and the right side equation (\ref{62}) has the
three solutions:
\begin{equation} \label{64}
\begin{array}{c}
\displaystyle{%
    z_{em}^2=\kappa x^2\varepsilon'(x), \qquad z_s^2=\frac{\beta_-(x)}{\kappa},\qquad z_0^2=-\frac{\beta_+(x)}{\kappa}. %
}
\end{array}
\end{equation}
The first two solutions $z_{em}^2, z_{s}^2$ describe, respectively,
the propagating electromagnetic and sound waves. The third solution
$z_{0}^2$ does not correspond to the propagating waves and will not
be considered.

Equation (\ref{62}) allows to obtain the corrections to the
solutions $z_{em}^2, z_{s}^2$, caused by the mutual influence of the
sound and electromagnetic waves. The solutions of (\ref{62}) are
sought in the form $z^2=z_{em}^2+\xi_{em}$ or $z^2=z_{s}^2+\xi_{s}$,
where $\xi_{em}\ll z_{em}^2$ and $\xi_{s}\ll z_{s}^2$. Then the
corrections to the spectrum of the electromagnetic and sound waves
will be determined by equations
\begin{equation} \label{65}
\begin{array}{c}
\displaystyle{%
    \xi_{em}^2+\frac{\zeta_+\zeta_-}{\big(\zeta_++\zeta_-\big)}\xi_{em}-\eta_B\frac{\Theta_{em}(x)}{\big(\zeta_++\zeta_-\big)}=0, %
}
\end{array}
\end{equation}
\begin{equation} \label{66}
\begin{array}{c}
\displaystyle{%
    \xi_{s}^2+\frac{\zeta_-\big(\zeta_--\zeta_+\big)}{\big(\zeta_+-2\zeta_-\big)}\xi_{s}-\eta_B\frac{\Theta_{s}(x)}{\big(\zeta_+-2\zeta_-\big)}=0, %
}
\end{array}
\end{equation}
where the designations are introduced
\begin{equation} \label{67}
\begin{array}{c}
\displaystyle{%
    \zeta_-\equiv\zeta_-(x)\equiv\kappa x^2\varepsilon'(x)-\frac{\beta_-(x)}{\kappa},\qquad %
    \zeta_+\equiv\zeta_+(x)\equiv\kappa x^2\varepsilon'(x)+\frac{\beta_+(x)}{\kappa}, %
}
\end{array}
\end{equation}
and also
\begin{equation} \label{68}
\begin{array}{c}
\displaystyle{%
    \Theta_{em}(x)\equiv-\kappa^{-1}\Omega\big(z_{em}^2,x\big)=\frac{1}{3}\kappa x^4\varepsilon'(x)\big(\varepsilon'(x)+2\big)\sin^2\theta, %
}
\end{array}
\end{equation}
\begin{equation} \label{69}
\begin{array}{c}
\displaystyle{%
    \Theta_{s}(x)\equiv-\kappa^{-1}\Omega\big(z_{s}^2,x\big)=\frac{\beta_-(x)}{\kappa}\Bigg\{
    x^2\frac{\big(\varepsilon'(x)+2\big)}{3}+\cos^2\theta\left[
    \frac{\beta_-(x)}{\kappa^2}\left(\frac{2}{3}-\frac{5}{3}\frac{1}{\varepsilon'(x)}\right)-x^2\big(\varepsilon'(x)-1\big)\right]
    \!\!\Bigg\}.
}
\end{array}
\end{equation}
Despite the assumed small value of the corrections $\xi_{em}$ and
$\xi_{s}$, as will be seen, it is not sufficient to restrict
ourselves to the linear in these quantities terms in equations
(\ref{65}) and (\ref{66}). First we consider separately the case of
propagation of waves without damping, and then take into account the
influence of the dissipative effects.

\section{Electromagnetic and sound waves in BEC without account of damping  }\vspace{-0mm}
The dispersion dependencies of the electromagnetic $z_{em}^2(x)$
(curve $A_1OC_2$) and sound $z_{s}^2(x)$ (curve $A_2OC_1$) waves
without account of their hybridization are presented graphically in
Fig.\,1. These curves always intersect at some frequency $x_*<1$. At
this frequency the function $\zeta_-(x_*)=\kappa
x^2\varepsilon'(x_*)-\kappa^{-1}\beta_-(x_*)=0$, and thus near $x_*$
in the linear approximation the required small corrections tend to
infinity, so that it becomes necessary to take into account the
quadratic terms in (\ref{65}) and (\ref{66}). Neglecting the
damping, when $\gamma=\tilde{\nu}=0,$ the corrections to the
dispersion laws of waves $z_{em}^2(x)$, $z_{s}^2(x)$ are determined
by the equations (\ref{65}) and (\ref{66}). The permittivity is real
in this case
\begin{equation} \label{70}
\begin{array}{c}
\displaystyle{%
    \varepsilon'(x)\equiv\varepsilon(x)=1+\frac{\big(\varepsilon_0-1\big)}{\big(1-x^2\big)}, %
}
\end{array}
\end{equation}
and tends to infinity at approaching from the side of low
frequencies to the resonance frequency $\omega_0$. In the frequency
range $\omega_0<\omega<\sqrt{\varepsilon_0}\omega_0$ the
permittivity is negative, and electromagnetic waves cannot propagate
in such medium. At $\omega>\sqrt{\varepsilon_0}\omega_0$ a
possibility of propagating of the electromagnetic waves appears
again, but we are not going to consider them in this high-frequency
region. The coordinates of the point of intersection of the
dispersion curves $x_*, k_*$ in neglect of hybridization (point $O$
in Fig.\,1) are determined by the equations
\begin{equation} \label{71}
\begin{array}{c}
\displaystyle{%
    \frac{\beta_-(x_*)}{x_*^2\varepsilon(x_*)}=\kappa^2\equiv\left(\frac{k_1}{k_2}\right)^{\!2},\qquad %
    k_*^2=k_1^2x_*^2\varepsilon(x_*)=k_2^2\beta_-(x_*).
}
\end{array}
\end{equation}
In the point of intersection
\begin{equation} \label{72}
\begin{array}{c}
\displaystyle{%
    \Theta_{em}(x_*)=\Theta_{s}(x_*)\equiv\Theta(x_*)=\frac{\kappa^2x_*^4\varepsilon(x_*)\big(\varepsilon(x_*)+2\big)}{3\big(\beta_-(x_*)+\beta_+(x_*)\big)}\sin^2\theta. %
}
\end{array}
\end{equation}
The account of the relation between the electromagnetic and sound
waves results in that at the frequency $x_*=\omega_*/\omega_0$ there
appear the jumps in the both branches determined by the equations
(\ref{65}) and (\ref{66}). The magnitudes of these jumps are equal
for both branches, but have the different signs. For the
electromagnetic wave:
\begin{equation} \label{73}
\begin{array}{c}
\displaystyle{%
    k_{em}^2\big(x_*+0\big)=k_1k_2\sqrt{\eta_B\Theta(x_*)}, \qquad %
    k_{em}^2\big(x_*-0\big)=-k_1k_2\sqrt{\eta_B\Theta(x_*)}, %
}
\end{array}
\end{equation}
and for the sound wave:
\begin{equation} \label{74}
\begin{array}{c}
\displaystyle{%
    k_{s}^2\big(x_*+0\big)=-k_1k_2\sqrt{\eta_B\Theta(x_*)}, \qquad %
    k_{s}^2\big(x_*-0\big)=k_1k_2\sqrt{\eta_B\Theta(x_*)}. %
}
\end{array}
\end{equation}
Due to hybridization the two continuous dispersion curves appear
($A_1B_1C_1$ and $A_2B_2C_2$ in Fig.\,1).

In the waves, propagation of which is described by these curves,
generally speaking, both the density of the condensate and the
electromagnetic field oscillate, but the amplitudes of the
oscillations prove to be comparable only near the frequency
$x_*=\omega_*/\omega_0$. The region  $A_1B_1$ of the first curve
mainly corresponds to the electromagnetic waves, which in the region
$B_1C_1$ transform into the sound waves. The region $A_2B_2$ of the
second curve mainly corresponds to the sound waves, which in the
region $B_2C_2$ transform into the electromagnetic waves. Note that
Fig.\,1 and other figures in the paper are given with an
illustrative purpose, so the parameters at which the calculations
are done are chosen from the illustrative considerations and can
essentially differ from the parameters of real systems.

It should be pointed out the important role of the dipole-dipole
interaction near the point of intersection of curves. Without
account of the dipole-dipole interaction the function (\ref{61}) has
the form
\begin{equation} \label{75}
\begin{array}{c}
\displaystyle{%
   \tilde{ \Omega}\big(z^2,x\big)=-\kappa
   z^2\left\{x^2-\frac{z^2}{\kappa\varepsilon(x)}\cos^2\theta\right\}.
}
\end{array}
\end{equation}
In contrast to the function (\ref{61}), which accounts for the
dipole-dipole interaction, the function (\ref{75}) contains only the
inverse permittivity giving a small contribution but does not
contain the permittivity in the first power which is large near the
resonant frequency. In the point of intersection of spectrums $x_*$
the function (\ref{75}) takes the value $\tilde{
\Omega}\big(z_*^2,x_*\big)=-\kappa^2x_*^4\varepsilon(x_*)\sin^2\theta$,
and the function (\ref{61}), accounting for the dipole-dipole
interaction, is
$\Omega\big(z_*^2,x_*\big)=(-1/3)\kappa^2x_*^4\varepsilon(x_*)\big(\varepsilon(x_*)+2\big)\sin^2\theta$.
Their ratio is
\begin{equation} \label{76}
\begin{array}{c}
\displaystyle{%
   \frac{\tilde{
\Omega}\big(z_*^2,x_*\big)}{\Omega\big(z_*^2,x_*\big)}=\frac{3}{\big(\varepsilon(x_*)+2\big)}
}
\end{array}
\end{equation}
small, since $\varepsilon(x_*)\gg 1$. Thus, taking into account of
the dipole-dipole interaction leads to a substantial increase of the
``repulsion'' of the branches of the spectrum.

\newpage
\begin{figure}[h!]
\vspace{-6mm} \hspace{-0mm}
\includegraphics[width = 8.0cm]{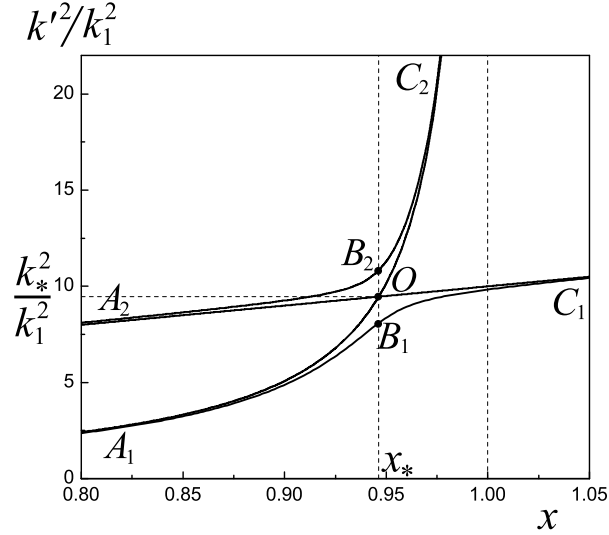} 
\vspace{-5mm}
\caption{\label{fig01} %
Dispersion curves without account of damping. %
The curve $A_1OC_2$ is the electromagnetic wave
$k'^2\big/k_1^2=x^2\varepsilon'(x)$, the curve $A_2OC_1$ is the
sound wave $k'^2\big/k_1^2=\beta_-(x)\big/\kappa^2$. %
The dispersion curves $A_1B_1C_1$ and $A_2B_2C_2$ arise as a result of hybridization. %
The calculation is carried out with an illustrative purpose with the parameters: %
$\kappa\equiv k_1\big/k_2=\sqrt{0.1}, \phi=10^{-3}, \varepsilon_0=2,
\theta=\pi/2, \eta_B=0.1, \gamma=0.1, \tilde{\nu}=0.$
}%
\end{figure}

\vspace{-2mm} The relation between the amplitudes of oscillations of
the BEC density $\delta n=\psi_0\delta\Psi=\tilde{n}e^{iQ({\bf
r},t)}$ ($\tilde{n}\equiv\psi_0\Psi_0$ is the amplitude of
oscillations of the density) and the electromagnetic field for the
curves which arose due to hybridization can be obtained from the
equations (\ref{27}). The dependencies of the ratio
$R\equiv\big|\big(\tilde{n}\big/n_0\big)\big/\big({\bf
\underline{b}}\cdot{\bf L}_1\big)\big|$ (where ${\bf \underline{b}}$
is the dimensionless amplitude of oscillations of the electric
field, and $\displaystyle{{\bf
L}_1\equiv-\frac{\big(\varepsilon(\omega)+2\big)}{3}\,{\bf
e}}+\big(\varepsilon(\omega)-1\big)\big({\bf e}\cdot{\bf
s}\big)\,{\bf s}$) for the branches $A_1B_1C_1$ and $A_2B_2C_2$ are
shown in Fig.\,2. As seen, the most strong coupling between the
electromagnetic and sound oscillations arises near the frequency $x_*$, %
where the sound wave is transformed into the electromagnetic wave
and vice versa.
\begin{figure}[h!]
\vspace{0mm} \hspace{-0mm}
\includegraphics[width = 8.0cm]{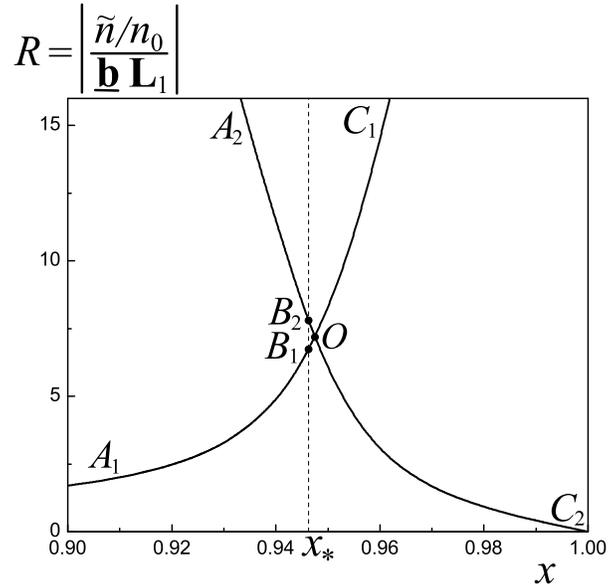} 
\vspace{-5mm}
\caption{\label{fig02} %
The ratio of the amplitude of oscillations of the BEC density to the
amplitude of oscillations of the electric field. %
The curves $A_1B_1C_1$ and $A_2B_2C_2$ correspond to the respective
branches in Fig.\,1. The calculation is carried out with the
parameters: $\kappa=\sqrt{0.1}, \phi=10^{-3}, \varepsilon_0=2,
\theta=\pi/2, \eta_B=0.1, \tilde{\nu}=0$.
}%
\end{figure}

In the long-wave limit one obtains the following dispersion law of
the sound waves
\begin{equation} \label{77}
\begin{array}{c}
\displaystyle{%
   \omega^2=\left(1-\frac{\hbar\omega_0}{2mc_B^2}\eta_Bf_D(\theta)\right)c_B^2k^2,
}
\end{array}
\end{equation}
so that the speed of a sound wave in BEC with account of the
dipole-dipole interaction is given by the relation
\begin{equation} \label{78}
\begin{array}{c}
\displaystyle{%
   c_D^2=\left(1-\frac{\hbar\omega_0}{2mc_B^2}\eta_Bf_D(\theta)\right)\!c_B^2.
}
\end{array}
\end{equation}
At the angle $\theta_0$, determined by formula the
$\cos^2\theta_0=1/3$, the function $f_D(\theta_0)=0$. In the angle
intervals $0<\theta<\theta_0$ and $\pi-\theta_0<\theta<\pi$ the
function $f_D(\theta)$ is negative, so that here the dipole-dipole
interaction leads to an increase of the speed of sound waves in
comparison with the Bogolyubov speed. In the angle interval
$\theta_0<\theta<\pi-\theta_0$, vice versa, the speed decreases
reaching its minimal value
$c_{D,min}^2=c_B^2\big(1-\eta_B\hbar\omega_0\big/6mc_B^2\big)$ at
$\theta=\pi/2$. Formula (\ref{78}) coincides with the formula for
the speed of propagation of sound in a system of Bose particles
without an intrinsic dipole moment, which is placed in a constant
electric field $E_0$, when the dipole moment $d_0=\alpha_0E_0$ is
induced by this field [29].

\section{Influence of dissipation on propagation of waves in BEC}\vspace{-0mm}
When the dissipation is taken into account, the real part of the
permittivity $\varepsilon'(x)$ is determined by formula (\ref{56}).
Its value remains finite near the resonance frequency
$x_0$. Thus, there are two possibilities: %
a) for a small damping rate of the electromagnetic wave, just as in
the non-dissipative case, the sound and electromagnetic dispersion
curves intersect in neglect of hybridization; %
b) the dispersion curve of the electromagnetic wave lies below the
sound branch, so that the intersection of branches is absent. %
Let us consider both possibilities.

\begin{figure}[b!]
\vspace{0mm} \hspace{-0mm}
\includegraphics[width = 8.0cm]{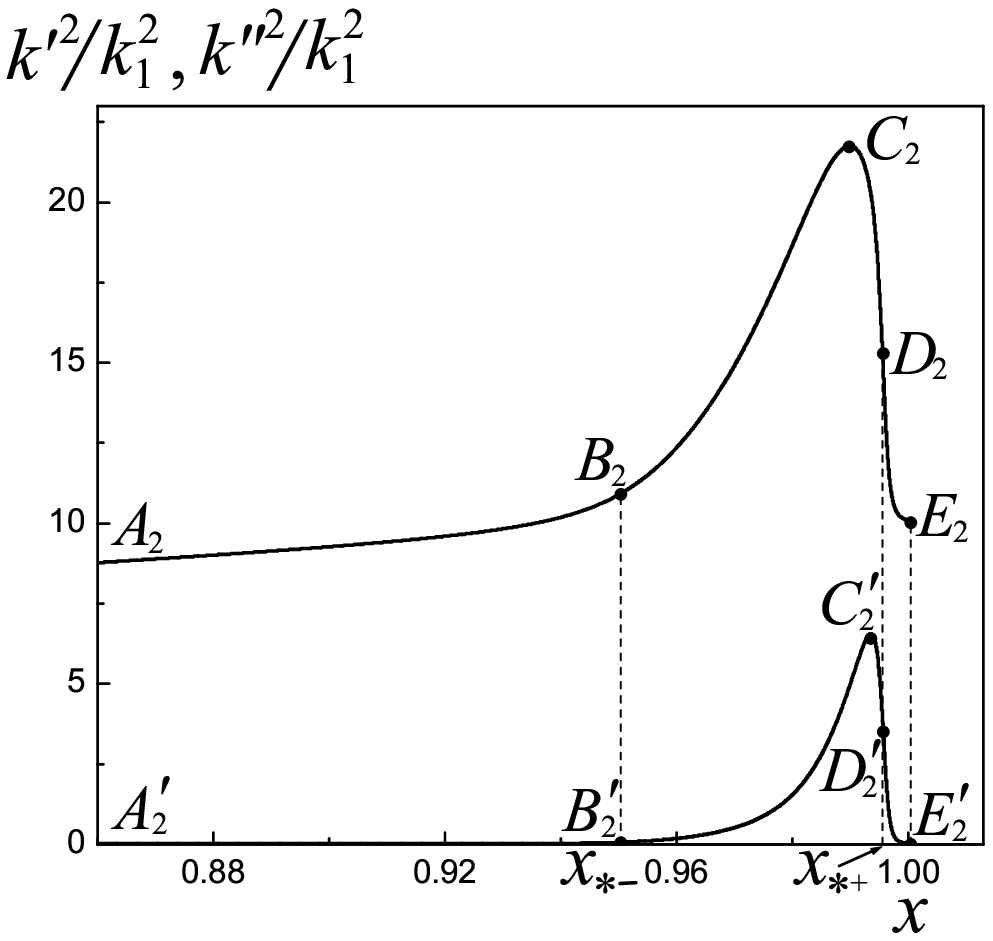} 
\vspace{-4mm}
\caption{\label{fig02} %
The curve $A_2B_2C_2D_2E_2$ $\big(k'^2\!\big/k_1^2\big)$ corresponds
to the branch $A_2B_2C_2$ for the non-dissipation case in Fig.\,1. %
The curve $A_2'B_2'C_2'D_2'E_2'$ $\big(k''^2\!\big/k_1^2\big)$ is
the damping on this branch. %
The calculation is carried out with the parameters: %
$\kappa=\sqrt{0.1}, \phi=10^{-3}, \varepsilon_0=2, \theta=\pi/2,
\eta_B=0.1, \gamma=0.1, \tilde{\nu}=3\cdot 10^{-2}.$
}%
\end{figure}

When one accounts for the damping, the points of the possible
intersection of the electromagnetic and sound dispersion curves are
determined by the solutions of the equation
\begin{equation} \label{79}
\begin{array}{c}
\displaystyle{%
   \varepsilon'(x_*)-\frac{\beta_-(x_*)}{\kappa^2x_*^2}=0,
}
\end{array}
\end{equation}
which has the solutions if the condition is satisfied
\begin{equation} \label{80}
\begin{array}{c}
\displaystyle{%
   \tilde{\nu}<\frac{\big(\varepsilon_0-1\big)}{2}\kappa^2,\qquad
   \frac{\nu}{\omega_0}<\frac{\big(\varepsilon_0-1\big)}{4}\frac{h\omega_0}{mc^2}.
}
\end{array}
\end{equation}
The equation (\ref{79}) under the condition (\ref{80}) has two
solutions. The smaller root $x_{*-}$ corresponds to the intersection
of the sound wave with the electromagnetic wave in the region of
normal dispersion, and the greater root $x_{*+}$ -- in the region of
anomalous dispersion. The dependencies of the squares of the real
and imaginary parts of the wave numbers on the frequency for two
branches, calculated by formulas (\ref{58}), (\ref{63}), (\ref{65}),
(\ref{66}), are shown in Figs.\,3,\,4. As seen, accounting for
dissipation results in that both dispersion curves become
nonmonotonous and in the region of anomalous dispersion the damping
of waves substantially increases.
\begin{figure}[t!]
\vspace{0mm} \hspace{-0mm}
\includegraphics[width = 8.0cm]{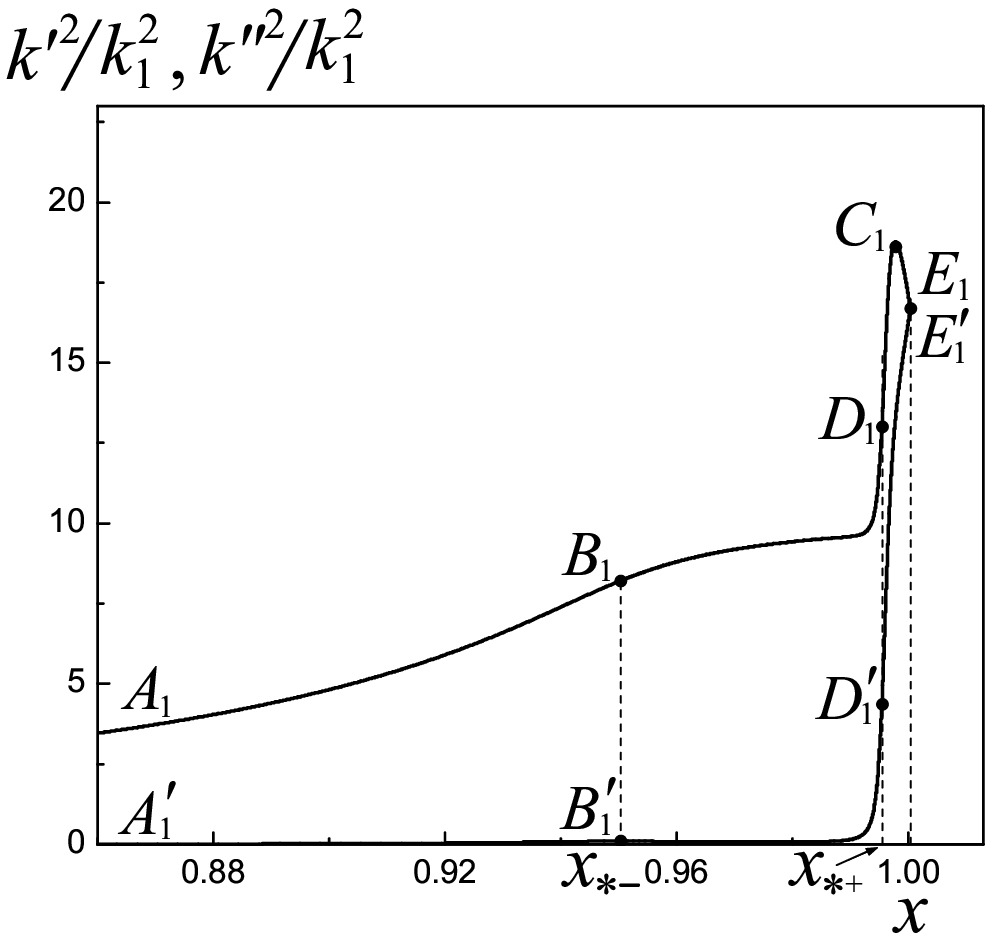} 
\vspace{-4mm}
\caption{\label{fig04} %
The curve $A_1B_1C_1D_1E_1$ $\big(k'^2\!\big/k_1^2\big)$ corresponds
to the branch $A_1B_1C_1$ for the non-dissipation case in Fig.\,1. %
The curve $A_1'B_1'D_1'E_1'$ $\big(k''^2\!\big/k_1^2\big)$ is the
damping on this branch. The calculation is carried out with the
parameters: $\kappa=\sqrt{0.1}, \phi=10^{-3}, \varepsilon_0=2,
\theta=\pi/2, \eta_B=0.1, \gamma=0.1, \tilde{\nu}=3\cdot 10^{-2}.$
}%
\end{figure}
\begin{figure}[b!]
\vspace{0mm} \hspace{-0mm}
\includegraphics[width = 8.0cm]{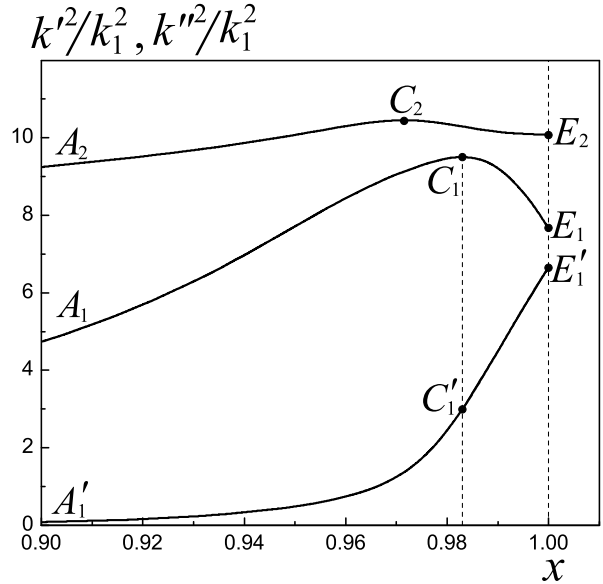} 
\vspace{-4mm}
\caption{\label{fig05} %
Dispersion curves for the sound and electromagnetic waves in the
absence of intersection of spectrums. The curve $A_1C_1E_1$
$\big(k'^2\!\big/k_1^2\big)$ is the dispersion law of the
electromagnetic waves, and the curve $A_1'C_1'E_1'$
$\big(k''^2\!\big/k_1^2\big)$ is their damping. The curve
$A_2C_2E_2$ $\big(k'^2\!\big/k_1^2\big)$ is the dispersion law of
the sound waves, which damping is small for the chosen parameters.
The calculation is carried out with the parameters:
$\kappa=\sqrt{0.1}, \phi=10^{-3}, \varepsilon_0=2, \theta=\pi/2,
\eta_B=0.1, \gamma=0.1, \tilde{\nu}=7\cdot 10^{-2}.$
}%
\end{figure}

If the condition (\ref{80}) is not fulfilled, then the equation
(\ref{79}) does not have a solution and the sound and
electromagnetic branches do not intersect. The dispersion curves and
the damping curve for the electromagnetic wave in this case are
shown in Fig.\,5. Here the mutual influence of the sound and
electromagnetic oscillations manifests itself in the appearance of
maximums in the points $C_1$ and $C_2$ on the dispersion curves
$A_1E_1C_1$ and $A_2E_2C_2$. Thus, even in the absence of
intersection of the electromagnetic and sound dispersion curves,
near the resonance frequency there exists their considerable
``interaction''.

\section{Conclusion}\vspace{-0mm}
In the paper it is theoretically studied the propagation of the
electromagnetic waves and sound oscillations in a Bose-Einstein
condensate of particles with an intrinsic dipole moment. At that
also the electric polarization of atoms under the action of the
constant electric field was taken into consideration. Bose-Einstein
condensate is described by the modified Gross-Pitaevskii equation
which accounts for the relaxation effects by means of introducing a
phenomenological dissipative coefficient connected to the third
viscosity coefficient and the time of uniform relaxation of the
particle number density in the condensate. The interaction of an
atom of the condensate with the electric field is accounted for in
the dipole approximation. The medium under study is anisotropic,
because there is a preferential direction, along which the dipole
moments of atoms are oriented. The tensor of permittivity of BEC, in
which atoms interact by means of the short-range forces and the
dipole-dipole interaction, is obtained. It is shown that such medium
possesses both the time and the spatial dispersion.

In the absence of dissipation or at a small dissipation the
dispersion curves of the electromagnetic and sound waves in neglect
of the interaction intersect at a frequency which is somewhat less
than the resonance frequency of an atom. Since atoms possess an
intrinsic dipole moment, there arises hybridization of the
electromagnetic and sound branches near the frequency of
intersection of the dispersion curves. Here the sound wave
transforms into the electromagnetic wave and vice versa, and the
oscillations of the condensate density are accompanied with the
oscillations of the electric field. It is obtained that taking into
account of the dipole-dipole interaction leads to a substantial
increase of the ``repulsion'' of the branches of the excitation
spectrum. It is shown that the account for the dissipation
conditioned by the relaxation of the macroscopic wave function of
the condensate and by the imaginary part of the permittivity, leads
to the appearance of regions with the anomalous dispersion.


In this work we considered hybridization of the sound and
electromagnetic waves at high frequencies close to the resonance
frequency of an atom. In the experiment [10] there was found a
resonant absorption of the microwave radiation in the superfluid
helium at a frequency close to 180\,GHz, which in [21] was
interpreted as an indication of the existence of the elementary
excitations with the energy gap, which exist together with the sound
excitations. Note also that the branch of excitations with the
energy gap can exist, for example, in the condensate of atoms and
their two-atom bound states [30]. Taking into account of the pair
correlations in a Bose system also leads to the appearance of a new
branch of excitations with the energy gap [31]. In the presence of
excitations with a gap the dispersion curve of these excitations
always intersects the dispersion curve of the electromagnetic waves
and there arises a phenomenon of the strong hybridization of
spectrums and the resonant absorption of radiation at more low
frequencies [21]. These effects also can be studied by the same
method that was used in this work.

\end{document}